\documentclass[11pt,twoside]{article}
\usepackage{./asp2014}

\aspSuppressVolSlug
\resetcounters

\newcommand\degd{\ifmmode^{\circ}\!\!\!.\,\else$^{\circ}\!\!\!.\,$\fi}

\newcommand{\msun}{{\rm\ M_{\sun}}}

\newcommand{\lsim}{\stackrel{\scriptstyle <}{\scriptstyle \sim}}
\newcommand{\gsim}{\stackrel{\scriptstyle >}{\scriptstyle \sim}}

\newcommand{\sgra}{Sgr~A*}

\newcommand{\ngvla}{ngVLA}

\begin{document}

\title{Galactic Center Pulsars with the \ngvla}
\author{Geoffrey C. Bower$^1$,  Shami Chatterjee$^2$, Jim Cordes$^2$, Paul Demorest$^3$, Julia S. Deneva$^4$, Jason Dexter$^5$, Michael Kramer$^6$, Joseph Lazio$^7$,  Scott Ransom$^8$,    Lijing Shao$^6$,
Norbert Wex$^6$, Robert Wharton$^6$
\affil{$^1$Academia Sinica Institute of Astronomy and Astrophysics, Hilo, HI 96720, USA; 
\email{gbower@asiaa.sinica.edu.tw}}
\affil{$^2$Department of Astronomy, Cornell University, Ithaca, NY 14853, USA;
\email{shami@astro.cornell.edu,cordes@astro.cornell.edu}}
\affil{$^3$NRAO, Socorro, NM, 87801, USA
\email{pdemores@nrao.edu}}
\affil{$^4$George Mason University, resident at the Naval Research Laboratory, Washington, DC 20575, USA; 
\email{iulia.deneva.ctr@nrl.navy.mil}}
\affil{$^5$Max Planck Institute for Extraterrestrial Physics, Garching, Germany;
\email{jdexter@mpe.mpg.de}}
\affil{$^6$Max-Planck-Institut f\"{u}r Radioastronomie, Auf dem H\"ugel 69, D-53121 Bonn, Germany;
\email{mkramer@mpifr-bonn.mpg.de,lshao@mpifr-bonn.mpg.de,wex@mpifr-bonn.mpg.de,wharton@mpifr-bonn.mpg.de}}
\affil{$^7$Jet Propulsion Laboratory, California Institute of Technology, CA 91109, USA
\email{Joseph.Lazio@jpl.nasa.gov}}
\affil{$^8$NRAO, Charlottesville, VA 22903, USA;
\email{sransom@nrao.edu}}
}


\paperauthor{Geoffrey C. Bower}{gbower@asiaa.sinica.edu.tw}{0000-0003-4056-9982}{Academia Sinica Institute of Astronomy and Astrophysics}{}{Hilo}{HI}{96720}{USA}
\paperauthor{Shami Chatterjee}{shami@astro.cornell.edu}{}{Cornell University}{Department of Astronomy}{Ithaca}{NY}{14853}{USA}
\paperauthor{Jim Cordes}{cordes@astro.cornell.edu}{}{Department of Astronomy}{Cornell University}{Ithaca}{NY}{14853}{USA}
\paperauthor{Paul Demorest}{pdemores@nrao.edu}{}{NRAO}{}{Socorro}{NM}{87801}{USA}
\paperauthor{Julia Deneva}{julia.deneva@gmail.com}{}{}{}{}{}{}{USA}
\paperauthor{Jason Dexter}{jdexter@mpe.mpg.de}{}{Max Planck Institute for Extraterrestrial Physics}{}{Garching}{}{}{Germany}
\paperauthor{Michael Kramer}{mkramer@mpifr-bonn.mpg.de}{}{Max Planck Institute f\"{u}r Radioastronomie}{}{Bonn}{}{}{Germany}
\paperauthor{Joseph Lazio}{Joseph.Lazio@jpl.nasa.gov}{}{Jet Propulsion Laboratory}{California Institute of Technology}{Pasadena}{CA}{91109}{USA}
\paperauthor{Scott Ransom}{sransom@nrao.edu}{}{NRAO}{}{Charlottesville}{VA}{22903}{USA}
\paperauthor{Lijing Shao}{lshao@mpifr-bonn.mpg.de}{}{Max Planck Institute f\"{u}r Radioastronomie}{}{Bonn}{}{}{Germany}
\paperauthor{Norbert Wex}{wex@mpifr-bonn.mpg.de}{}{Max Planck Institute f\"{u}r Radioastronomie}{}{Bonn}{}{}{Germany}
\paperauthor{Robert Wharton}{wharton@mpifr-bonn.mpg.de}{}{Max Planck Institute f\"{u}r Radioastronomie}{}{Bonn}{}{}{Germany}

\paperauthor{Sample~Author3}{Author3Email@email.edu}{ORCID_Or_Blank}{Author3 Institution}{Author3 Department}{City}{State/Province}{Postal Code}{Country}


\vspace{0.5cm}
Pulsars in the Galactic Center (GC) are important probes of General Relativity, 
star formation, stellar dynamics, stellar evolution, and the 
interstellar medium.  Despite years of searching, only a handful of pulsars
in the central $0.5^\circ$ are known.  
The high-frequency sensitivity of \ngvla\ will open a new window for discovery and characterization of pulsars in the GC.  A pulsar in orbit around the GC black hole, \sgra, will provide an unprecedented probe of black hole physics and General Relativity.

\section{Scientific Goals}

Currently, only six pulsars in the central $30^\prime$ of the Galaxy have 
been detected.  The most spectacular of these, J1745$-$2900, is a transient
magnetar located only 0.1~pc in projection from \sgra.  This small 
number stands in sharp contrast to predictions for the number of pulsars 
based on the rapid star formation rate and high density of young massive
stars that can serve as pulsar progenitors.  Numerous searches at
a wide range of wavelengths have been carried out.

Until recently,
the standard explanation for the absence of pulsars
was the presence of hyperstrong interstellar scattering that smeared
pulses over a timescale of $>10^2$ s at low frequencies.  Higher frequency
searches reduced scattering effects but suffered from lower sensitivity
due to the typical steep spectrum of pulsars.  
The discovery of J1745$-$2900 in~2013, however, suggested that scattering
effects from the hyperstrong screen cannot fully account for the 
absence of detected pulsars, leading to the ``missing pulsar'' problem.
Possible solutions to the absence of observed pulsars include more complex
scattering models, stellar population synthesis arguments, and 
mechanisms for the suppression of the pulsar emission mechanism.

The population of 
millisecond pulsars (MSPs), which are the most prized targets for dynamical studies,
has not been probed at all by existing observations.  
Even with the much reduced scattering strength inferred by J1745$-$2900,
nearly all MSPs would remain undetected by existing surveys through
a combination of time smearing and low flux density.
\ngvla\ Galactic Center (GC) pulsar surveys will probe both the slow pulsar and MSP populations.

The \ngvla\ will make dramatic changes to our understanding of the GC pulsar
population and potentially lead to the discovery of a pulsar in a short-period
bound orbit to \sgra.  This will come primarily through the factor of~10
improvement in sensitivity at high frequencies that the \ngvla\ promises.

Specific science goals are
\begin{itemize}

\item{Searching for and Timing a Pulsar Bound to \sgra:  It has been
demonstrated that pulsars orbiting \sgra\ will be superb probes for
studying the properties of the central supermassive black hole.
It is sufficient to find and time a normal, slowly rotating pulsar in a
reasonable orbit, in order to measure the mass of \sgra\ with a
precision of $1M_\odot$(!), to test the cosmic censorship conjecture to
a precision of about~0.1\% and to test the no-hair theorem to a
precision of~1\%.  These tests are possible even with a rather modest timing
precision of 100~$\mu$s due to the large mass of \sgra\ and the
measurement of relativistic and classical spin-orbit coupling, including
the detection of frame-dragging. }

\item{Searching for and Characterizing the GC Pulsar Population:  Finding GC pulsars will not only lead to unique
studies of the General Relativistic description of black holes,
but we also gain invaluable information about the GC region itself: the characteristic age distribution of the
discovered pulsars will give insight into the star formation history;
MSPs can be used as accelerometers to probe the local
gravitational potential; the measured dispersion and scattering measures
(and their variability) would allow us to probe the distribution,
clumpiness and other properties of the central interstellar medium; this
includes measurements of the central magnetic field using Faraday
rotation. Proper motions of young pulsars can be used to point back to
regions of recent star formation and/or supernova remnants.
Broadly, we define the GC pulsar population to be that which falls within the Central Molecular Zone (diameter $\sim 250$~pc).
}

\item{Characterizing the Scattering Environment: Interstellar scattering 
prevents detection of pulsars at low frequencies.  This scattering appears to
be strongly variable as a function of position towards the \hbox{GC}.  Observations
of low frequency masers, extragalactic background sources, as well as known
pulsars can characterize the scattering screen and provide insights into
the window function for pulsar detection.}
\end{itemize}

The proximity of the Galactic Center and the power of the \ngvla\ combine to provide an unprecedented and unique opportunity to study the environment of a galactic nucleus through pulsars.

\section{Pulsar Searching in the GC}

Numerous searches for GC pulsars have been carried out over a wide
range of frequencies.  Prior to~2013, these searches had led to the detection of only five pulsars in the central $0.5^\circ$ \citep[e.g.,][]{2000ASPC..202...37K,2006MNRAS.373L...6J,2009ApJ...702L.177D,2010ApJ...715..939M,2013IAUS..291...57S,2013IAUS..291..382E}.  Theoretical expectations 
based on the high star formation rate and the density of high mass
stars in the central molecular zone suggest that there should be hundreds
or thousands of detectable pulsars within this region \citep{2004ApJ...615..253P}.
In particular, the immediate environment of the \sgra\ contains a 
$10^5 \msun$ cluster of massive stars with ages $\sim 4$~Myr--8~Myr 
bound to the black hole, which are possible progenitors to neutron stars and pulsars \citep{2003ApJ...597..323B,2011ApJ...741..108P}.

\begin{figure}
\includegraphics[width=0.5\textwidth]{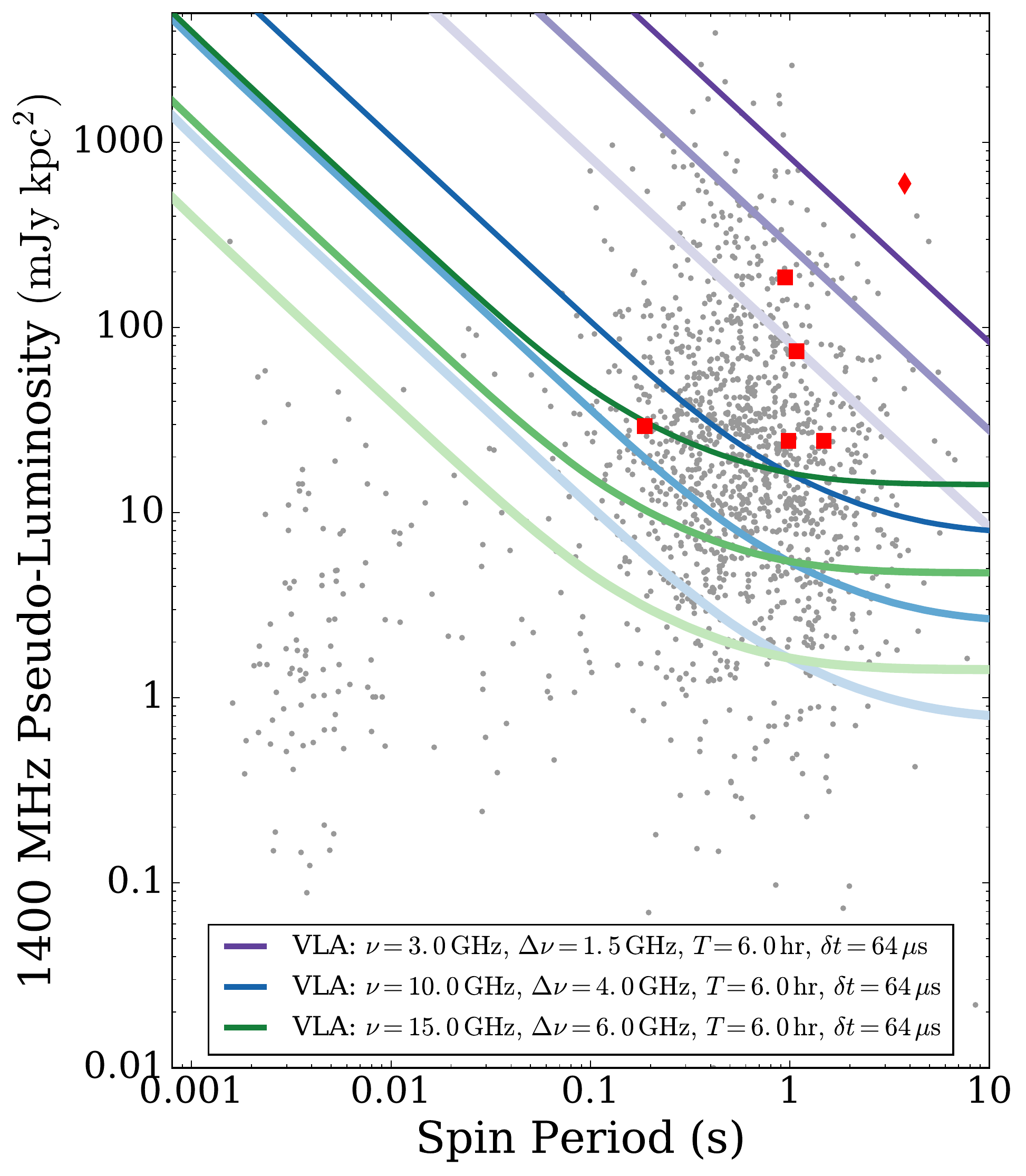}
\includegraphics[width=0.5\textwidth]{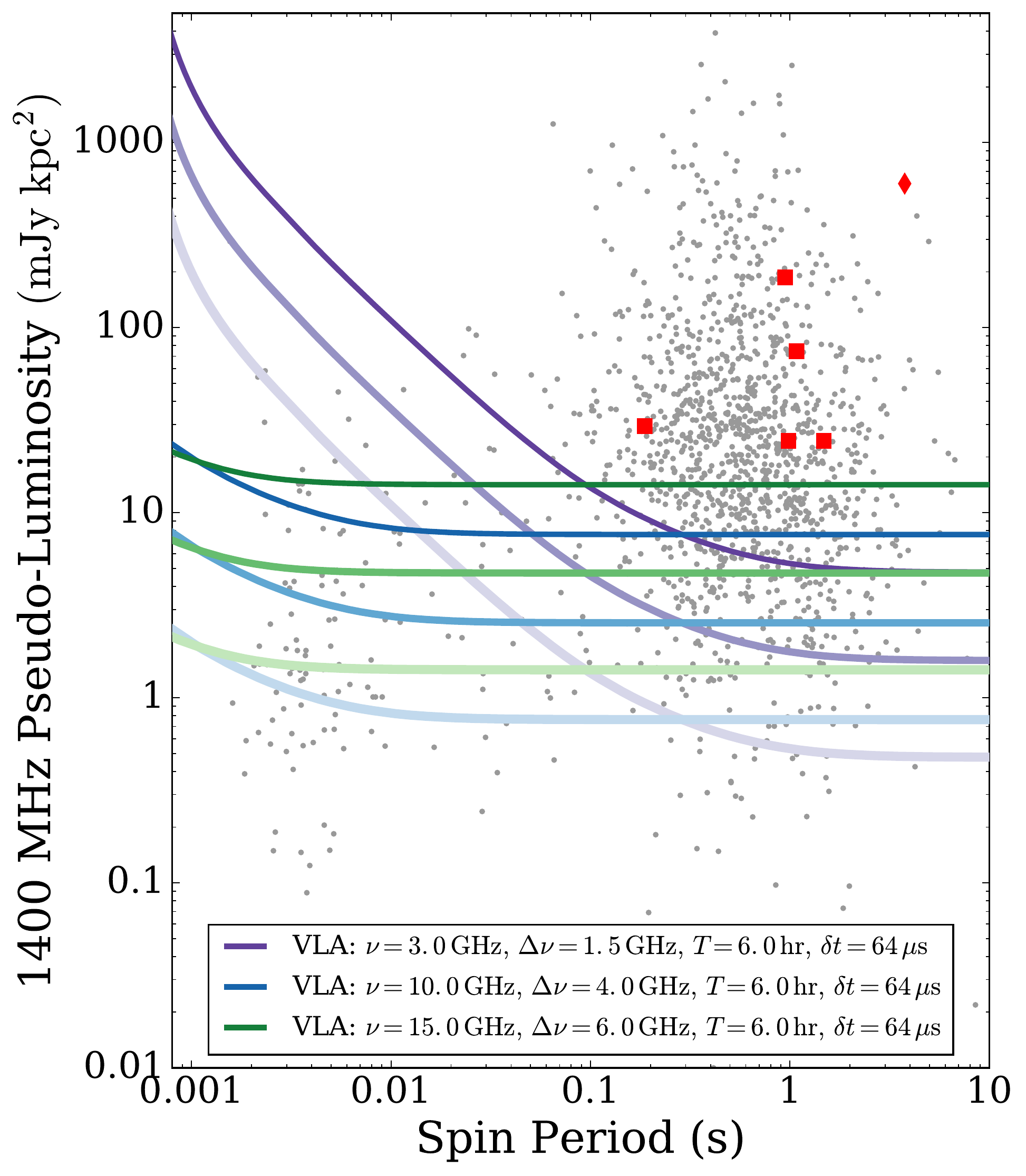}
\caption{Field and GC pulsars (grey dots and red symbols, respectively)
in period-pseudo-luminosity space for hyperstrong scattering (left) and magnetar-like scattering (right).  The red diamond indicates the GC magnetar J1745$-$2900.  Curves indicate sensitivity
of the 1, 3, and $10 \times$ the VLA at three frequencies (darker to lighter curves).  Sensitivity curves include nominal VLA and \ngvla\ system performance and scaling for average pulsar spectral index.
\label{fig:periodluminosity}
}
\end{figure}

The strong interstellar scattering towards the GC may account for
the absence of detected pulsars, especially at low radio frequencies.
Turbulent electrons along the line of sight will produce angular
broadening of sources in the image domain and temporal broadening of
sources in the time domain.  
The temporal scattering scales for a thin scattering screen is related
to the angular broadening  by the following equation:
\begin{equation}
\tau_s = 6.3 {\rm\ s\ } \times {\left( D \over 8.5 {\rm kpc} \right)} {\left( \theta_s \over 1.3 {\rm\ arcsec} \right)^2} \left( {D \over \Delta} - 1 \right) \nu^{-4},
\label{eqn:tau}
\end{equation}
where $D = 8.3 \pm 0.3$~kpc \citep{2010RvMP...82.3121G}
is the distance to the \hbox{GC}, $\Delta$ is the distance from the GC to the scattering medium,
and $\nu$ is the observing frequency
in GHz \citep{1997ApJ...475..557C}.  The observed angular diameter~$\theta_s$ is extrapolated to a frequency of 1~GHz using a scaling of $\nu^{-2}$.  
For Kolmogorov turbulence with an inner scale $r_{in} > b_{max}$,
we expect $\theta_s \propto \nu^{-2}$ and $\tau_s \propto \nu^{-4}$.
Several pieces of evidence demonstrate that scattering may be important
in the \hbox{GC}.  \sgra\ itself is observed to show substantial 
angular broadening, as are OH masers and some extragalactic background
sources.  These pieces of evidence have been used
to argue for a hyperstrong scattering screen that is located within
the central few hundred parsecs of the GC \citep{1998ApJ...505..715L}.  If the source to screen distance is small, an observed angular
broadening implies a large temporal broadening.  In the case of 
the hyperstrong model, $\tau_s \gsim 10^3$~s at~1~GHz, sufficient
to make even slow pulsars undetectable up to $\sim 10$~GHz.

The discovery and characterization of the GC magnetar J1745$-$2900 has indicated that the hyperstrong scattering model cannot hold for all
of the GC \citep{2013Natur.501..391E,2014ApJ...780L...3S,2014ApJ...780L...2B}.  The GC magnetar, with $P=3.76$~s is detected to frequencies 
as low as $\sim 1$~GHz, with a characteristic $\tau_s=1.3$~s at~1~GHz.
The GC magnetar also shows angular broadening 
identical to that of \sgra. Together, these results imply a source to screen 
distance of order 6~kpc, well outside of the \hbox{GC}, and the ability to 
readily detect pulsars in the GC at frequencies of a few GHz.  
It remains unclear whether the magnetar-like scattering screen describes
all of the GC region or may be patchy.  Characterization of the
other GC pulsars show distances to the scattering screen $\lsim 2$~kpc
\citep{2017MNRAS.471.3563D}.  
Other environmental effects may also alter the neutron star population in the GC \citep{2014ApJ...783L...7D}.

In the case of both hyperstrong and magnetar-like scattering, there is
still a substantial benefit in reduction of scattering that is achieved
by going to higher frequencies.  This benefit has to be offset against 
the steep spectral index of both slow pulsars and MSPs and 
decreasing sensitivity of many search telescopes with increasing
frequency.  The average slow pulsar has a spectral index of $\alpha\approx -1.4$
\citep{2013MNRAS.431.1352B} although some objects including magnetars have flat spectra to very high frequencies \citep{2017MNRAS.465..242T}.

Figure~\ref{fig:periodluminosity} shows the distribution of known field pulsars along with sensitivity curves for the VLA and the ngVLA at
a range of frequencies and for hyperstrong and magnetar-like scattering.  These curves reveal that for magnetar-like scattering the \ngvla\ will have sensitivity to discover the majority of field pulsars at the GC and a significant fraction of MSPs.  In the more challenging case of hyperstrong scattering, the high frequency \ngvla\ is essential for obtaining sensitivity to slow pulsars.

\begin{figure}[tb]
  \centering
\includegraphics[width=0.95\textwidth]{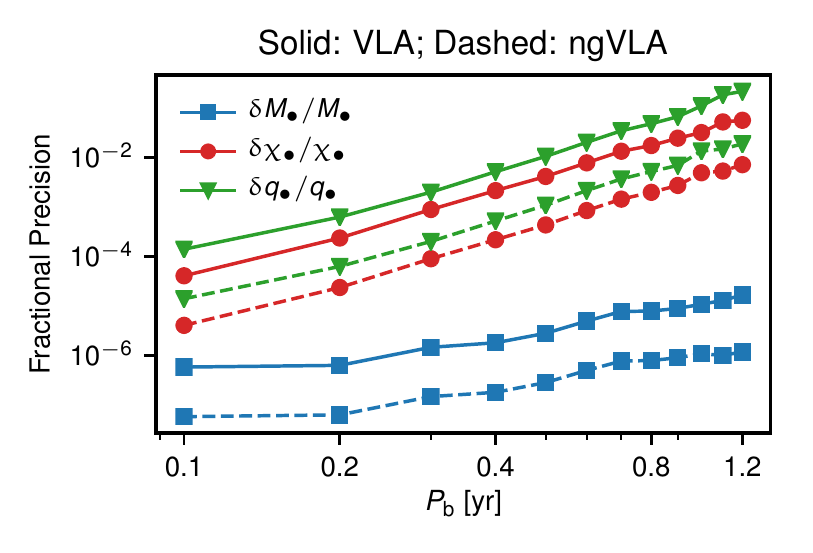}
\vspace*{-3ex}
\caption{%
Constraints on black hole parameters from timing observations
of a pulsar in a bound orbit for the VLA (solid lines) and for the 
\ngvla\ (dashed lines).    Constraints are given for black hole 
mass ($M_\bullet$), 
spin ($\chi_\bullet$), and 
quadrupole ($q_\bullet$).  These 
constraints were calculated for
a $P=0.5$~s pulsar with orbital eccentricity $e=0.8$ and 
timing residuals of $\sigma_{\rm TOA}=1$~ms and~0.1~ms for the VLA 
and \ngvla, respectively.
Simulations include weekly observations over a 5~yr interval.
\label{fig:bhparams}
}
\end{figure}

\section{Constraints on Black Hole Parameters}


In order to estimate the impact of ngVLA on timing of a pulsar around Sgr A* and testing the no-hair theorem, we have extended the work by \cite{wk99} and \cite{lwk+12}. We have set up a numerical integration scheme for orbits around \sgra, using MCMC parameter estimation code;  we extract the parameter uncertainties and covariances.  Figure~\ref{fig:bhparams} shows estimates of 
the uncertainty in black hole mass, spin, and quadrupole moment for a set
of pulsar orbits timed with the VLA and the \hbox{ngVLA}.  
Results for  a more complete set of orbits will be presented in a future
publication (Shao et al., in prep.).  Effectively, the precision on these parameters in this regime scales directly with increased collecting area.

\begin{figure}
\includegraphics[width=0.5\textwidth]{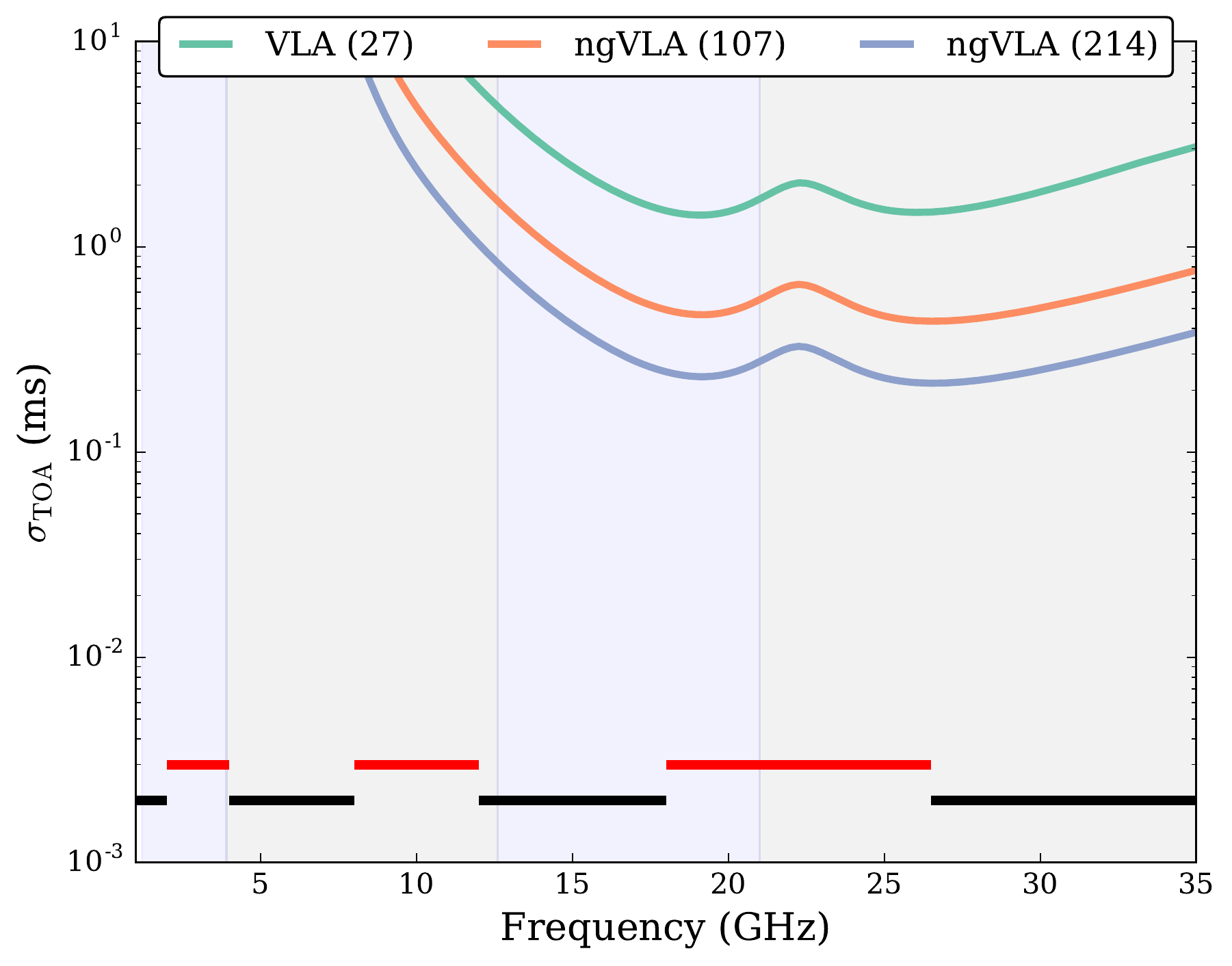}
\includegraphics[width=0.5\textwidth]{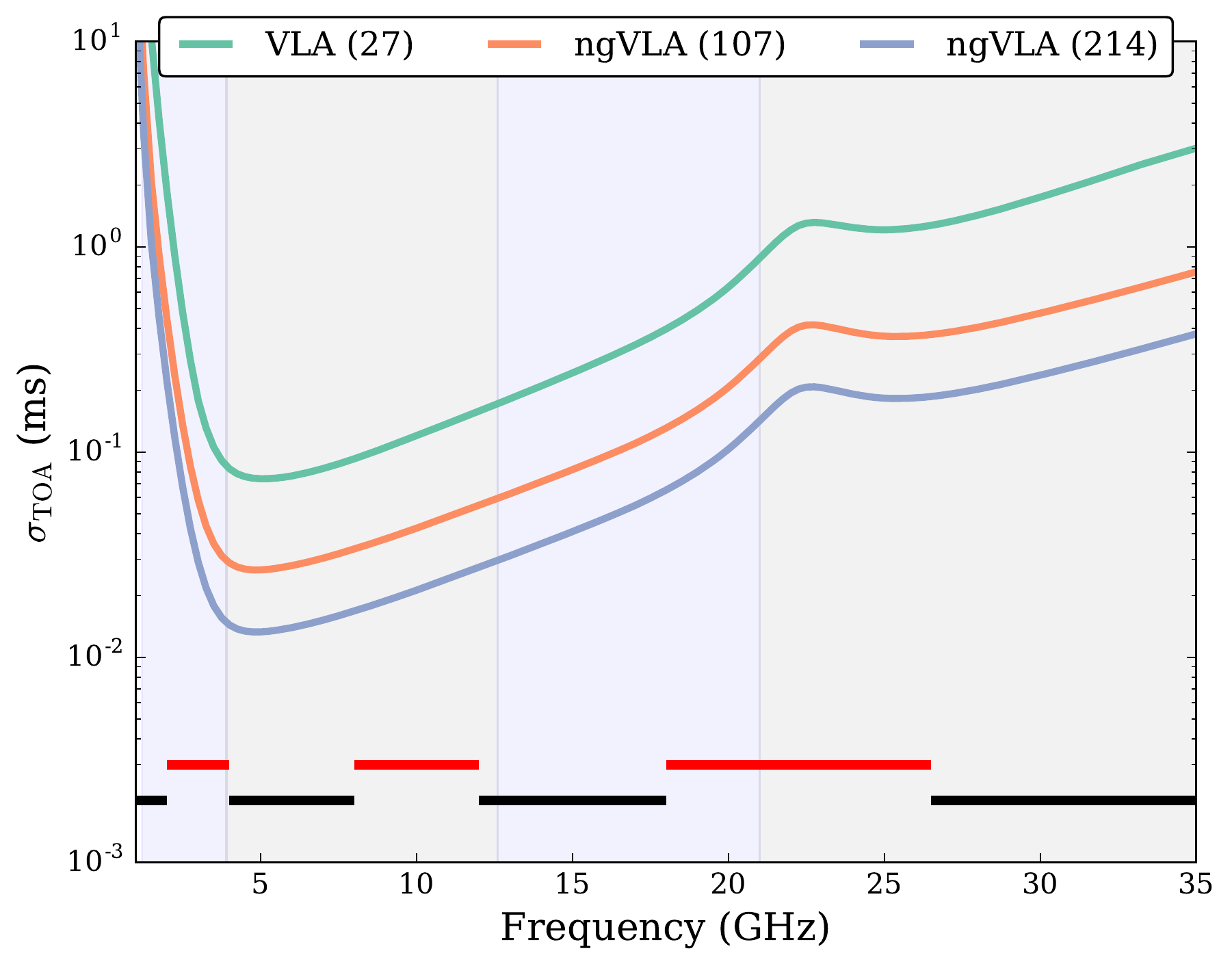}
\caption{TOA residuals for a $P=0.5$~s pulsar with a 10~ms pulsar width,
using an integration time of~4~hr and a bandwidth of~1~GHz.  The residuals are shown for hyperstrong scattering with $\tau=2300$~s at~1~GHz (\textit{left}) and for a level of scattering comparable to the magnetar J1745$-$2900, with $\tau=1.3$~s at~1~GHz (\textit{right}).
\label{fig:toa}
}
\end{figure}

The parameter constraints are governed primarily by the rms time of
arrival (TOA) residuals,
$\sigma_{\rm TOA}$.  For given intrinsic pulsar emission properties and scattering
properties, $\sigma_{\rm TOA}$ scales inversely with collecting area.  Figure~\ref{fig:toa} shows estimates of the timing error for two scattering models
(hyperstrong and magnetar) as a function of frequency for the VLA, \ngvla\
with~107 antennas, and \ngvla\ with~214 antennas.

General Relativistic effects are best measured in systems where the observational duration is greater than the orbital period, but constraints on General Relativity and BH properties can still be obtained for longer period systems.  We consider the case of a pulsar with a 15~yr period observed over a 5~yr interval during pericenter.  The BH mass can be constrained with \ngvla\ to an accuracy $\sim 10^{-5}$.  Lense-Thirring precession parameters will be found with an signal-to-noise ratio between~10 and~1000, depending on the details of orbit and the observations.  At large radii, perturbations to the pulsar orbit through stellar interactions will limit the accuracy with which General Relativity constraints can be made.  Next generation adaptive optics on large optical telescopes imaging the stellar cluster may enable modeling of these perturbations.

Another limitation of current GC pulsar searches is the insensitivity 
to a large fraction of binary pulsars.  The minimum detectable pulsar 
pseudo-luminosity ($L = S d^2$) for a pulsar survey scales roughly as 
\begin{equation}
   L_{\rm min} \propto d^{-2} G^{-1} T_{\rm obs}^{-1/2}
\end{equation}
where $d$ is distance, $G$ is the telescope gain, and $T_{\rm obs}$ is 
the observing time.  Because the GC is so far away, GC searches have 
typically observed for as long as possible ($\approx 6$~hr at the VLA).  
While this increases the sensitivity to faint isolated pulsars, it can 
also reduce the effectiveness of typical binary pulsar search methods.  
For example, the acceleration search technique is only effective for 
orbital periods with $P_{\rm orb} \gtrsim 10 T_{\rm obs}$ 
\citep{jk91,rce03}.  A VLA GC pulsar search that observes for 6~hr would be most sensitive to binary pulsars with orbits longer than about~60~hr, which includes only about~50\% of known pulsar binaries (and 
only 30\% of binaries in globular clusters).

The increased sensitivity of the ngVLA means that flux density limits 
achieved by the VLA can be reached in much shorter observing times, thus 
allowing for the possibility of detecting much shorter period binary 
pulsars.  For $(G_{\rm ng} / G_{\rm VLA}) = 5$, the ngVLA could reach 
the same flux density limit as a 6~hr VLA observation in only  
$T_{\rm ng} = T_{\rm VLA}/\sqrt{5} \approx 15~{\rm min}$.  In this shorter 
observing span, an acceleration search could be used to find pulsars 
with orbits as short as 2.5~hr, which would include about 95\% of all 
known pulsar binaries. 

Long baseline astrometry of the pulsar relative to \sgra\ can also contribute to parameter constraints for long-period systems.  For the case of~100~$\mu$as astrometric error \citep{2015ApJ...798..120B}, we can measure the orientation of the pulsar orbit in the sky, i.e., the longitude of the ascending node with sufficient precision to improve General Relativistic constraints.

\section{Other Probes of Black Hole Physics and the Uniqueness of \ngvla}

There are a number of experiments currently seeking to characterize \sgra, the
properties of the \hbox{SMBH}, and General Relativity.  These include the Event
Horizon Telescope (EHT) and its millimeter wavelength imaging of \sgra\ and large optical telescope campaigns to measure stellar 
orbits near \sgra.  \ngvla\ timing of a 
pulsar in a bound orbit compares favorably against these techniques and provides important complementary information \citep{pwk16}.

EHT constraints will primarily arise from modeling the static and time-domain images of the immediate region surrounding \sgra\ out to a radius of $\sim 10 R_{\mathrm{S}}$
\citep{2014ApJ...784....7B}.  General Relativity and BH properties introduce image distortions that can be translated into parameter constraints.  These measurements, however, must be interpreted in light of the complex astrophysics of accretion, jet launching, and particle acceleration in the vicinity of the \hbox{BH}.  Simulations have shown that EHT results will be able to measure the BH mass, spin, and quadrupole moment.

Stellar astrometry with existing and future large optical telescopes has the potential to measure orbital precession.  Current constraints on the black hole mass from stellar orbits have an accuracy of order 10\% \citep[e.g.,][]{2010RvMP...82.3121G} and future measurements with existing facilities are projected to improve on these results substantially \citep{2018MNRAS.tmp..464W}.  The closest star, S2, has a period of~15~yr and is currently going through periastron passage.  Improved instruments may discover fainter stars in shorter orbits leading to greater precision.  Sub-milliarcsecond astrometry provides a length-scale precision that is orders of magnitude less precise than millisecond-accuracy pulsar timing, leading to complementary but correspondingly less accurate BH and General Relativity parameter constraints.  

\section{The Accretion Flow, ISM, and Stars}

Pulsed emission provides a powerful probe of the plasma and magnetic field along the line of sight through a variety of measurements.  The dispersion measure (DM) gives the total electron column density.  The rotation measure (RM) gives the integrated line of sight magnetic field and electron density.  The scattering measure (SM) characterizes turbulent plasma, which is typically localized to individual regions along the line of sight.

A short period pulsar will orbit inside of the Bondi radius ($\sim 10^5 R_S$; $P\sim 10^3$~yr) of \sgra.  Polarimetry of \sgra\ and the magnetar J1745$-$2900  has shown that a substantial fraction of the observed \sgra\ RM originates from the accretion flow on scales between~10~R${}_{\mathrm{S}}$ and $10^5$~R${}_{\mathrm{S}}$, providing strong constraints on the accretion rate and the nature of the radiatively inefficient accretion flow \citep{2003ApJ...588..331B,2007ApJ...654L..57M}.  A pulsar in an eccentric orbit would provide a measure of the radial profile of the accretion flow, inaccessible to any other technique.

On larger angular scales, a collection of GC pulsars would provide multiple probes of the interstellar medium (ISM) in the GC region \citep{2016MNRAS.459.3005S}.  With a sufficient density of pulsars, the \ngvla\ can map out magnetized and ionized structures throughout the central molecular zone.  Astrometry of these pulsars can provide origins of individual pulsars in known SNR or young stellar clusters.

\ngvla\ pulsar studies will complement \ngvla\ and SKA studies of the GC ISM through other means.  The scattering medium has been studied in the past through imaging of stellar masers and extragalactic background sources 
\citep{vld91,vlf92,fdc94,1998ApJS..118..201L}.   Sensitivity and angular resolution constraints have limited observations, however, to a small number of sight lines. The \ngvla\ will provide the 
sensitivity and resolution to measure the angular sizes of a wide range of maser species and a large number of background sources and map out the spatial variations of 
the scattering towards the inner degree of the \hbox{GC}.  This will provide 
important insight into both the source of scattering and characterization of
the scattering that can lead to optimization of pulsar searches.

\section{Implications for Array Design}

A number of elements of the array design are constrained by our scientific 
goals.

\begin{itemize}
\item{Correlator and beamformer requirements:  Searches for pulsars will make use of beamforming capabilities as well as high-time resolution imaging with the correlator.  Different regimes of the search space will be optimized through tradeoffs in field of view and bandwidth.  Timing observations will rely primarily on wideband receivers and beamformer backends.}

\item{Effect of the various receiver options on GC searches:  Pulsar
searching requires high instantaneous sensitivity that cannot always be
traded off against greater integration time.  High frequency ($\sim 30$~GHz) sensitivity
is likely to be required to avoid the effects of scattering for detection
of MSPs.}

\item{Effect of array configuration on GC searches:  Searching and
imaging primarily require a compact configuration.  Timing observations can be obtained with beamforming of extended configurations but these will introduce challenges in calibration and availability due to weather.  Characterization of
the scattering environment and pulsar astrometry will require $>1000$ km baselines to resolve compact
sources and achieve sub-milliarcsecond positional accuracy.}
\end{itemize}

\section{Summary}

The order of magnitude increase in high frequency sensitivity that the \ngvla\ provides will be essential for the discovery and characterization of pulsars in the Galactic Center, the nearest and best studied galactic nucleus.  Characterization of the  population within the Central Molecular Zone will give new insights into the ISM  and stellar populations of one of the most dynamic regions in the Galaxy. Discovery of a pulsar in orbit around \sgra\ will provide an unprecedented opportunity for fundamental physics in the environment of a black hole.  

\acknowledgements~\hbox{SC}, \hbox{JMC}, \hbox{PD}, \hbox{JL}, and~SR are members of the NANOGrav Physics Frontier Center, which is supported by the National Science Foundation award 1430284.  \hbox{MK}, \hbox{LS}, \hbox{NW}, and \hbox{RW} acknowledge financial support by the European Research Council (ERC) for the ERC Synergy Grant BlackHoleCam under contract no.~610058. JSD acknowledges support by NASA under grant DPR~S--15633--Y.
Part of this research was carried out at the Jet Propulsion Laboratory, California Institute of Technology, under a contract with the National Aeronautics and Space Administration.   




\end{document}